\documentclass[pra,aps,twocolumn,showpacs]{revtex4-1}

\usepackage[utf8]{inputenc}
\usepackage[T1]{fontenc}
\usepackage[english]{babel}

\usepackage{graphicx}

\usepackage{amsmath,amssymb}
\usepackage{bm}

\usepackage{siunitx}

\usepackage[version=3]{mhchem}

\usepackage{acronym}

\newcommand{\dd}{\ensuremath{\mathrm{d}}}

\newcommand{\ii}{\ensuremath{\mathrm{i}}}

\newcommand{\eff}{\ensuremath{\text{eff}}}

\newcommand{\final}{\ensuremath{\text{f}}}

\newcommand{\group}{\ensuremath{\text{g}}}

\newcommand{\tar}{\ensuremath{\text{tar}}}

\newcommand{\well}{\ensuremath{\text{w}}}

\newcommand{\numg}{\ensuremath{{N_{\text{G}}}}}
\newcommand{\numw}{\ensuremath{{N_{\text{w}}}}}

\newcommand{\calp}{\ensuremath{\mathcal{P}}}
\newcommand{\calt}{\ensuremath{\mathcal{T}}}

\newcommand{\pt}{\ensuremath{\mathcal{PT}}}

\newcommand{\complex}{\ensuremath{\mathbb{C}}}

\newcommand{\imag}{\ensuremath{\operatorname{Im}}}
\newcommand{\real}{\ensuremath{\operatorname{Re}}}

\newcommand{\abs}[1]{\ensuremath{\left| #1 \right|}}

\newcommand{\eqph}{\ensuremath{\phantom{{}=}}}

\newcommand{\firstpderiv}[2]{\ensuremath{\frac{\partial #1}{\partial  #2}}}

\newcommand{\firsttderiv}[1]{\ensuremath{\frac{\dd #1}{\dd t}}}

\newcommand{\mat}[1]{\ensuremath{\bm{\mathrm{#1}}}}
\newcommand{\name}[1]{\textsc{#1}}
\newcommand{\nn}{\nonumber}
\newcommand{\jt}{\ensuremath{\tilde{j}}}

\renewcommand{\vec}[1]{\ensuremath{\bm{#1}}}
\usepackage{hyperref}

\begin{document}

\newacro{BEC}{Bose-Einstein condensate}
\newacro{GPE}{Gross-Pitaevskii equation}

\title{Tilted optical lattices with defects as realizations of
  \texorpdfstring{$\pt$}{PT} symmetry in Bose-Einstein condensates}
\author{Manuel Kreibich}
\author{J\"org Main}
\author{Holger Cartarius}
\author{G\"unter Wunner}
\affiliation{Institut f\"ur Theoretische Physik 1, Universit\"at
  Stuttgart, 70550 Stuttgart, Germany}

\begin{abstract}
  A $\pt$-symmetric Bose-Einstein condensate can theoretically be
  described using a complex optical potential, however, the
  experimental realization of such an optical potential describing the
  coherent in- and outcoupling of particles is a nontrivial task. We
  propose an experiment for a quantum mechanical realization of a
  $\pt$-symmetric system, where the $\pt$-symmetric currents are
  implemented by an accelerating Bose-Einstein condensate in a titled
  optical lattice. A defect consisting of two wells at the same energy
  level then acts as a $\pt$-symmetric double-well if the tilt in the
  energy offsets of all further wells in the lattice is varied in
  time. We map the time-dependence of the amplitudes of a frozen
  Gaussian variational ansatz to a matrix model and increase the
  system size step by step starting with a six-well setup. In terms of
  this simple matrix model we derive conditions under which two wells
  of the Hermitian multi-well system behave \emph{exactly} as the two
  wells of the $\pt$-symmetric system.
\end{abstract}


\maketitle

\section{Introduction}
\label{sec:introduction}

One of the postulates of quantum mechanics tells us that an observable
physical quantity is represented by a Hermitian operator. The
postulate ensures that every eigenvalue of such an operator is real,
and, since every result of a measurement corresponds to an eigenvalue
of an operator, all measured quantities are real. One may now ask, if
there is the possibility that, when dealing with a non-Hermitian
Hamiltonian, an entirely or partially real eigenvalue spectrum
occurs. The answer brings us to the notion of $\pt$ symmetry.

$\pt$ symmetry stands for the combined action of the operators parity
reflection $\calp$ and time or motion reversal $\calt$. A
$\pt$-symmetric Hamiltonian does not necessarily have to be
Hermitian. Nevertheless, \name{Bender} and \name{Boettcher} found in
their paper from 1998 \cite{bender98} that there are parameter regimes
in $\pt$-symmetric systems where the eigenvalue spectra are entirely
real. The non-Hermiticity entering $\pt$-symmetric systems can be
interpreted in that this is an effective description of an open
quantum system with a situation of balanced gain and loss.

Due to a close analogy between the Schrödinger equation and the
equations describing the propagation of light in structured wave
guides, the theory of $\pt$ symmetry was applied to these optical
systems \cite{ruschhaupt05, klaiman08}. Soon, the first
$\pt$-symmetric optical systems could be realized \cite{guo09,
  rueter10}. Other analogies with quantum mechanics could be used to
create $\pt$-symmetric systems, including laser modes
\cite{chong11,ge11,liertzer12}, electronics
\cite{schindler11,ramezani12,schindler12}, and microwave cavities
\cite{bittner12}.

It was proposed \cite{klaiman08} that a system similar to
$\pt$-symmetric wave guides could be realized using a \ac{BEC} in a
double-well potential, where in one well particles are injected, while
from the other well particles are removed. However, as was already
mentioned by the authors of Ref.~\cite{klaiman08} care must be taken
with regard to the interaction in \acp{BEC} since this could modify or
even destroy the $\pt$-symmetric properties. Nevertheless, \acp{BEC}
could provide an experimental realization of $\pt$ symmetry in a
genuine quantum system, which up to date is still missing.

In a schematic approach, the $\pt$-symmetric Bose-Hubbard model for
two modes has been considered, and it could be shown that the
mean-field limit is a good approximation even in the presence of a
complex non-Hermitian potential
\cite{graefe08a,graefe08b,graefe10}. In terms of the \ac{GPE} and a
model $\delta$-type potential it was confirmed that the nonlinear term
entering the dynamical equation does not destroy the features of
$\pt$-symmetry, rather these properties are maintained and even new
prospects do appear \cite{cartarius12a, haag14}. The extension of the
model to a realistic external potential gives qualitatively the same
results \cite{dast13}.

In all those investigations the $\pt$-symmetric potential was
predetermined and no detailed mechanism was explained on how to
actually realize $\pt$ symmetry, which leaves some discomfort. As
mentioned above, a non-Hermitian system is an effective description of
an underlying Hermitian system, so when discussing in detail the
realization of $\pt$-symmetry, this Hermitian system must be given.

It is the purpose of this work to demonstrate that a $\pt$-symmetric
external potential for \acp{BEC} can be realized with tilted optical
lattices with a defect. The defect consisting of two potential wells
at the same height, i.\,e., a discontinuity in the tilt, acts as a
$\pt$-symmetric subsystem. The in- and outfluxes of the condensate's
probability density can be modeled as imaginary $\pt$-symmetric
contributions.

When such a system is explicitly given, an experimental realization of
$\pt$ symmetry in a \ac{BEC} is possible. This is an important step in
the research of $\pt$-symmetric \acp{BEC}, since even when no future
experiment is done according to the proposals of this work, the
investigations performed in the above mentioned references are
justified and the gap originating from the sketchiness of the
$\pt$-symmetric potential is closed.

\begin{figure}[t]
  \centering
  \includegraphics[width=0.49\linewidth]{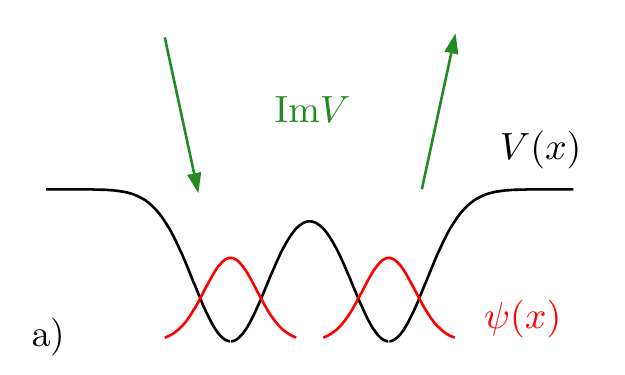}
  \includegraphics[width=0.49\linewidth]{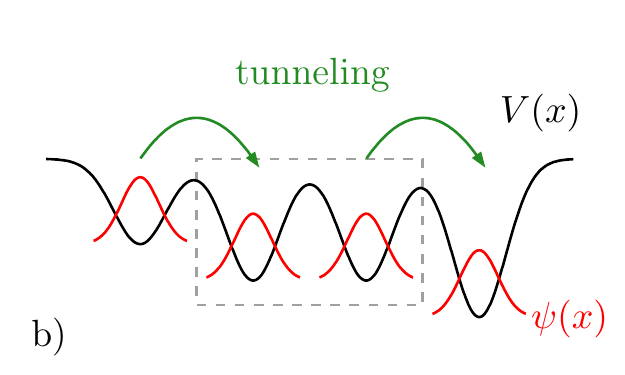}
  \includegraphics[width=0.49\linewidth]{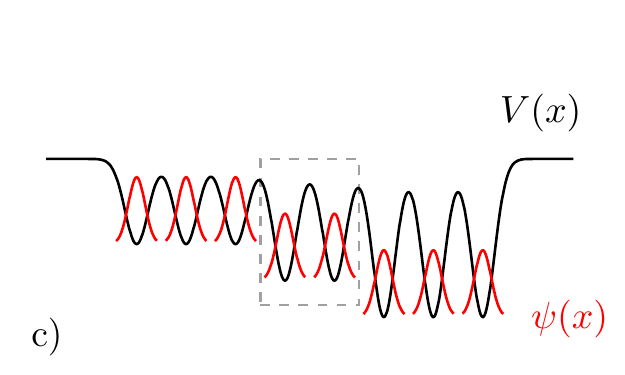}
  \includegraphics[width=0.49\linewidth]{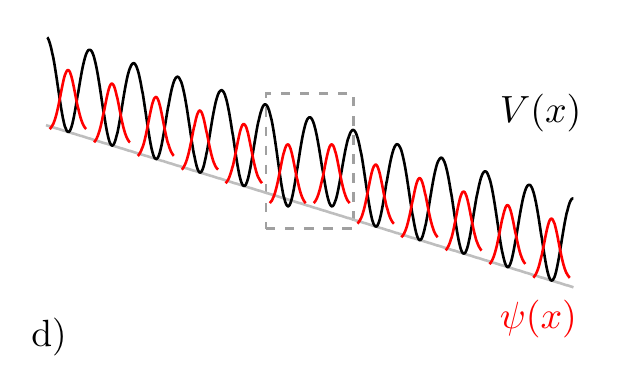}
  \caption{(Color online) (a) $\pt$-symmetric two-well system as
    discussed in Sec.~\ref{sec:matrix-models}. The imaginary part of
    the external potential induces currents from the environment to
    the left well and from the right well to the environment. (b) By
    embedding the two-well system in two additional outer wells the
    tunneling currents may be used as an implementation of
    $\pt$-symmetry \cite{kreibich14}. (c) With more outer wells the
    reservoir is larger, which can be advantageous. (d) In the limit
    of an infinite number of reservoir wells the system may be
    interpreted as the two-well system embedded in an optical
    lattice.}
  \label{fig:basic-idea}
\end{figure}

There are several possibilities for a realization of $\pt$
symmetry. In \cite{single14}, the authors investigate the pointwise
coupling of \acp{BEC} with the possible exchange of particles at those
points. In \cite{gutoehrlein15} another double-well system is
suggested as a particle reservoir for the implementation of
$\pt$-symmetry. However, in both references a coupling with no further
specification is used. It is not clear, how such a coupling can be
realized experimentally, which leaves some ambiguity in the possible
realizations.

By contrast, we follow an alternative approach, which gives us a
complete and realistic description of a possible realization. The
basic idea on how to realize a $\pt$-symmetric two-well system as
shown in Fig.~\ref{fig:basic-idea}(a) is sketched in
Fig.~\ref{fig:basic-idea}(b): In this simple case the two-well system
is embedded into a four-well system with two additional time-dependent
wells. It has already been shown that the inner wells of this system
can \emph{exactly} behave as a $\pt$-symmetric double-well
\cite{kreibich14}. In this setup the occurrence of the tunneling
currents from the outer wells to the inner ones may be used as a
realization of $\pt$-symmetric gain and loss effects. Thus, the
additional wells serve as particle reservoirs. Since there is a finite
number of particles in the reservoirs $\pt$ symmetry can only be
realized for a finite time period. In principle an arbitrary number of
particles could be filled in the reservoir. However, such a large
difference of the number of particles between the embedded and
reservoir wells could be a large challenge for an experiment, both for
preparing and measuring the particle numbers.

Hence, in this paper we extend our previous investigations and use
step by step more wells for a realization, see
Fig.~\ref{fig:basic-idea}(c). In principle, an arbitrary number of
wells, and thus an arbitrary number of particles in the reservoir can
be used, thus, maximizing the time that is available for realizing
$\pt$ symmetry. Using more wells consequently leads to the embedding
of two wells into an optical lattice, Fig.~\ref{fig:basic-idea}(d),
which allows us to use new tools and to obtain a new viewpoint on our
method. The experimental realization of a multi-well potential that is
necessary for our realization is possible with the method presented in
\cite{henderson09}.

This paper is organized as follows: In Sec.~\ref{sec:matrix-models} we
present the matrix model that is used throughout this work. In
Sec.~\ref{sec:real-pt-symm} we discuss how to realize $\pt$-symmetry
within that model and derive conditions that must be fulfilled for the
Hermitian system such that $\pt$-symmetry can be realized. In
Sec.~\ref{sec:results-simulations}, these results are applied to six
and more wells. Finally, in Sec.~\ref{sec:conclusion-outlook} we
summarize our work and give an outlook to future work.

\section{Matrix models}
\label{sec:matrix-models}

A \ac{BEC} is described in the limit of large particle numbers by the
\ac{GPE}
\begin{align}
  \label{eq:gpe}
  \ii \hbar \firstpderiv{}{t} \psi(\vec{r},t) = \left[ -
    \frac{\hbar^2}{2m} \Delta + V(\vec{r},t) + g
    \abs{\psi(\vec{r},t)}^2 \right] \psi(\vec{r},t),
\end{align}
where $g = 4 \pi \hbar^2 N a / m$ is the interaction strength with the
$s$-wave scattering length $a$ and the particle number $N$. The
external potential $V$ is time-dependent in this work and given by
\begin{align}
  \label{eq:gauss-pot}
  V(\vec{r},t) = \sum\limits_{k=1}^\numw V^k(t) \exp \left( -\frac{2
      x^2}{w_x^2} - \frac{2 y^2}{w_y^2} - \frac{2 [z -
      s_z^k(t)]^2}{w_z^2} \right),
\end{align}
which is a superposition of Gaussian profiles for $\numw$ wells. The
quantity $w_\alpha$ designates the width in the direction
$\alpha$. The time-dependence is due to the depths $V^k(t)<0$ of each
well and the displacements along the $z$-direction $s_z^k(t)$. Such a
potential can be created in an experiment with the method described in
\cite{henderson09}.

In \cite{kreibich14} we presented a method to map the
\ac{GPE}~\eqref{eq:gpe} with the continuous
potential~\eqref{eq:gauss-pot} to a discrete, finite-dimensional
system using a frozen Gaussian variational ansatz. This ansatz reads
\begin{align}
  \psi(\vec{r},t) = \sum\limits_{k=1}^\numg \psi^k(t) g^k(\vec{r}),
\end{align}
with the amplitudes $\psi^k$ and the $\numg$ Gaussian functions
\begin{align}
  g^k(\vec{r}) = \exp \left[ - A_x^k x^2 - A_y^k y^2 - A_z^k
    (z-q_z^k)^2 \right].
\end{align}
Throughout this paper we assume that $\numg = \numw$, i.\,e.\ we use
one Gaussian function located in one well, $q_z^k \approx
s_z^k$. After integrating out the spatial dependencies and performing
a symmetric orthogonalization \cite{lowdin50} we arrive at
\begin{align}
  \label{eq:dgpe}
  \ii \hbar \dot{\vec{\psi}} = \mat{H} \vec{\psi}
\end{align}
with the Hamiltonian in nearest-neighbor approximation
\begin{align}
  \label{eq:ham}
  H_{kl}(t)=
  \begin{cases}
    E_k(t) + g_k \abs{\psi_k(t)}^2, &\text{for } k=l, \\
    -J_{kl}(t), &\text{for } \abs{k-l}=1, \\
    0, &\text{otherwise}.
  \end{cases}
\end{align}
The time-dependencies of the parameters of the
potential~\eqref{eq:gauss-pot} are mapped to time-dependencies of the
onsite energies $E_k(t)$ and the tunneling elements $J_{kl}(t)$. The
relation between the parameters $V^k$, $s_z^k$ on one hand and $E_k$,
$J_{kl}$ on the other hand can be found in \cite{kreibich14}.

As an example, for a $\pt$-symmetric two-well system $\numw=2$ with
$\real V^1 = \real V^2$ and $\imag V^1 = -\imag V^2$ we obtain
\begin{align}
  \label{eq:ham2}
  \mat{H}=
  \begin{pmatrix}
    \ii \Gamma + g \abs{\psi_1}^2 & -J_{12} \\
    -J_{12} & -\ii \Gamma + g \abs{\psi_2}^2
  \end{pmatrix},
\end{align}
where for convenience the energy has been moved such that the onsite
energies are zero. A detailed and comprehensive analysis of this
system, which is sketched in Fig.~\ref{fig:basic-idea}(a), can be
found in \cite{graefe12}.

\section{Realizing \texorpdfstring{$\pt$}{PT} symmetry}
\label{sec:real-pt-symm}

\subsection{Preliminary calculations}
\label{sec:prel-calc}

Before going on, we present some calculations that are necessary for
later considerations in terms of the few-mode model. For a discrete
system with $\numw$ wells, a state is described by the vector
$\vec{\psi} \in \complex^\numw$. This time-dependent vector is a
solution of the time-dependent discrete \ac{GPE}~\eqref{eq:dgpe} with
the Hamiltonian~\eqref{eq:ham}. The modulus squared of vector entry
$k$ is interpreted as proportional to the number of particles in well
$k$, that is $n_k = \abs{\psi_k}^2$. When $\vec{\psi}$ is normalized
to unity for $N$ particles, then the number of particles in well $k$
is $N_k = N n_k$. The time derivative of $n_k$ can be calculated using
the \ac{GPE}, which yields
\begin{align}
  \label{eq:n-deriv}
  \firsttderiv{n_k} &= j_{k-1,k} - j_{k,k+1},
\end{align}
where we used $H_{kl} = H_{lk}^*$. The current from well $k$ to well
$l$ with $\abs{k-l}=1$, $j_{kl}$, is given by
\begin{align}
  j_{kl} = \frac{\ii}{\hbar} J_{kl} \left( \psi_k \psi_l^* - \psi_k^*
    \psi_l \right) = - j_{lk}.
\end{align}
For later convenience we define the ``modified'' current by
\begin{align}
  \jt_{kl} = \ii \left( \psi_k \psi_l^* - \psi_k^* \psi_l \right) =
  j_{kl} \frac{\hbar}{J_{kl}},
\end{align}
which is just the dimensionless current. The current $j_{kl}$ is only
defined for adjacent wells, whereas the modified current $\jt_{kl}$
can be used for arbitrary distances between the wells. The
time derivative of the modified current can be calculated by similar
means, this gives
\begin{align}
  \label{eq:jt-deriv}
  \hbar \firsttderiv{\jt_{kl}} &= \left( E_k + g_k n_k - E_l - g_l n_l
  \right) C_{kl} - \zeta_{kl},
\end{align}
where we defined the quantity
\begin{align}
  C_{kl} = \psi_k \psi_l^* + \psi_k^* \psi_l = C_{lk}
\end{align}
and the abbreviation
\begin{multline}
  \zeta_{kl} = J_{k-1,k} C_{k-1,l} + J_{k,k+1} C_{k+1,l} - J_{l-1,l}
  C_{k,l-1} \\ - J_{l,l+1} C_{k,l+1}.
\end{multline}
Finally, the time derivative of $C_{kl}$ is given by
\begin{align}
  \label{eq:c-deriv}
  \hbar \firsttderiv{C_{kl}} &= \left( E_l + g_l n_l - E_k - g_k n_k
  \right) \jt_{kl} + \eta_{kl}
\end{align}
with
\begin{multline}
  \eta_{kl} = J_{k-1,k} \jt_{k-1,l} + J_{k,k+1} \jt_{k+1,l} -
  J_{l-1,l} \jt_{k,l-1} \\ - J_{l,l+1} \jt_{k,l+1}.
\end{multline}

\subsection{Derivations of conditions}
\label{sec:deriv-cond}

The nonlinear $\pt$-symmetric two-mode model is given by the
Hamiltonian~\eqref{eq:ham2}. The time derivative of the number of
particles per well, $n_k = \abs{\psi_k}^2$, leads to the closed set of
coupled real differential equations
\begin{subequations}
  \label{eq:obs-two-mode}
  \begin{align}
    \firsttderiv{n_1} &= -j_{12} + 2 \Gamma n_1 / \hbar, \\
    \firsttderiv{n_2} &= j_{12} - 2 \Gamma n_2 / \hbar, \\
    \hbar \firsttderiv{\tilde{j}_{12}} &= 2 J_{12} (n_1-n_2) + g
    (n_1-n_2) C_{12}, \\
    \hbar \firsttderiv{C_{12}} &= - g (n_1-n_2) \tilde{j}_{12}.
  \end{align}
\end{subequations}
When we want to realize the system using a larger Hermitian system, we
have to make sure that two wells of the larger system behave exactly
as the two wells of the $\pt$-symmetric two-mode model, that means the
time derivatives of the observables must be equivalent to
Eqs.~\eqref{eq:obs-two-mode}.

We now think of embedding the two wells of the $\pt$-symmetric system
in a larger system with an arbitrary number of wells. The two wells
that shall behave like the $\pt$-symmetric two-mode model (which are
called \emph{embedded wells} from now on) have the indices $m$ and
$m+1$. The Hamiltonian for the multi-mode model is given in
nearest-neighbor approximation by Eq.~\eqref{eq:ham}. The onsite
energies $E_k$ and the tunneling elements $J_{kl}=J_{lk}$ are in
general time-dependent and real despite of the energies for the
embedded wells, which are given by $E_m=E_{m+1}=0$, and the tunneling
element $J_{m,m+1}$ that is time-independent.

We continue by calculating the time derivatives of the observables for
the embedded wells $m$ and $m+1$ of the multi-mode
model~\eqref{eq:ham}, which yields
\begin{subequations}
  \label{eq:obs-multi-mode}
  \begin{align}
    \firsttderiv{n_m} &= j_{m-1,m} - j_{m,m+1}, \\
    \firsttderiv{n_{m+1}} &= j_{m,m+1} - j_{m+1,m+2}, \\
    \hbar \firsttderiv{\jt_{m,m+1}} &= 2 J_{m,m+1} \left( n_m -
      n_{m+1} \right) \nn\\
    &\eqph + \left( g_m n_m - g_{m+1} n_{m+1} \right)
    C_{m,m+1} \nn\\
    &\eqph - J_{m-1,m} C_{m-1,m+1} + J_{m+1,m+2} C_{m,m+2}, \\
    \hbar \firsttderiv{C_{m,m+1}} &= - \left( g_m n_m - g_{m+1}
      n_{m+1} \right) \jt_{m,m+1} \nn\\
    &\eqph + J_{m-1,m} \jt_{m-1,m+1} - J_{m+1,m+2} \jt_{m,m+2}.
  \end{align}
\end{subequations}
If we now identify the wells $1$ and $2$ of the $\pt$-symmetric
two-mode model~\eqref{eq:obs-two-mode} with wells $m$ and $m+1$ of the
general multi-mode model~\eqref{eq:obs-multi-mode}, we find that the
nonlinear interaction strengths must be equal for the multi-mode
model, i.\,e.\ $g_m=g_{m+1}=g$. Then, we obtain conditions that must
be fulfilled such that the embedded wells of the multi-mode model
behave exactly as the two wells of the $\pt$-symmetric system. These
conditions are
\begin{subequations}
  \label{eq:conditions}
  \begin{align}
    \label{eq:conditions-1} j_{m-1,m} &= 2 \Gamma n_m / \hbar, \\
    \label{eq:conditions-2} j_{m+1,m+2} &= 2 \Gamma n_{m+1} / \hbar, \\
    \label{eq:conditions-3} J_{m-1,m} C_{m-1,m+1} &= J_{m+1,m+2}
    C_{m,m+2}, \\
    \label{eq:conditions-4} J_{m-1,m} \jt_{m-1,m+1} &=
    J_{m+1,m+2} \jt_{m,m+2}.
  \end{align}
\end{subequations}
This is the first central result on our way to realize $\pt$ symmetry:
If these conditions can be fulfilled for a finite time period, then in
this time we are able to realize $\pt$ symmetry. It remains to show
that these conditions can indeed be fulfilled by giving the matrix
elements of the multi-mode model~\eqref{eq:ham}, and further that this
choice can be established for a finite time.

Before we move on, we discuss the conditions~\eqref{eq:conditions}
considering the number of independent constraints. With four
independent equations we must use four matrix elements to fulfill
these equations. It can be shown that only three conditions are
independent and the fourth can be deduced \cite{kreibich14}. Thus, we
treat Eqs.~\eqref{eq:conditions-1}--\eqref{eq:conditions-3} as
independent.

\subsection{Solution for matrix elements}
\label{sec:solut-matr-elem}

So far, we have derived conditions under which the embedded wells in a
Hermitian multi-well model behave exactly as the $\pt$-symmetric
two-mode model. Condition~\eqref{eq:conditions-3} can simply be
fulfilled by setting the tunneling elements
\begin{align}
  \label{eq:j-sol}
  J_{m-1,m} &= d C_{m,m+2}, & J_{m+1,m+2} &= d C_{m-1,m+1},
\end{align}
with an arbitrary time-dependent function $d=d(t)$. There are two
conditions remaining, namely
\begin{subequations}
  \label{eq:two-cond-n}
  \begin{align}
    \hbar j_{m-1,m} &= J_{m-1,m} \jt_{m-1,m} \stackrel{!}{=} 2 \Gamma
    n_m, \\
    \hbar j_{m+1,m+2} &= J_{m+1,m+2} \jt_{m+1,m+2} \stackrel{!}{=} 2
    \Gamma n_{m+1}.
  \end{align}
\end{subequations}
Since both tunneling elements are determined by Eqs.~\eqref{eq:j-sol},
there are no free parameters left to fulfill
Eqs.~\eqref{eq:two-cond-n} at a given time.

However, there is a possibility to ensure that
Eqs.~\eqref{eq:two-cond-n} are fulfilled. To see this, we take the
time derivative of these equations, which gives
\begin{subequations}
  \label{two-cond-rewrite}
  \begin{align}
    2 \firsttderiv{\Gamma} n_m + 2 \Gamma \firsttderiv{n_m} &=
    \firsttderiv{J_{m-1,m}} \jt_{m-1,m} \nn\\
    &\eqph + J_{m-1,m} \firsttderiv{\jt_{m-1,m}}, \\
    2 \firsttderiv{\Gamma} n_{m+1} + 2 \Gamma \firsttderiv{n_{m+1}} &=
    \firsttderiv{J_{m+1,m+2}} \jt_{m+1,m+2} \nn\\
    &\eqph + J_{m+1,m+2} \firsttderiv{\jt_{m+1,m+2}},
  \end{align}
\end{subequations}
where we allow for a time-dependent value of $\Gamma$. For further
evaluation we need to calculate the time derivatives of the quantities
$\jt_{kl}$, which are given in Eq.~\eqref{eq:jt-deriv}.

For the time derivatives of the tunneling elements we need the
time derivative of $d$ in Eq.~\eqref{eq:j-sol}, which is not
determined at this point. We assume that $\dot{d}$ contains terms that
are at most linear in the energies $E_{m-1}$ and $E_{m+2}$. As it
turns out later, this is indeed the case for all choices of the
time-dependent function $d(t)$ in this work. We write
\begin{align}
  \label{eq:d-val-def}
  \hbar \dot{d} = D_{m-1} E_{m-1} + D_{m+2} E_{m+2} + D,
\end{align}
where $D$ contains all other quantities and all $D$-quantities are
independent of $E_{m-1}$ and $E_{m+2}$.

When inserting all calculated expressions,
Eqs.~\eqref{two-cond-rewrite} give a linear system of equations for
the onsite energies $E_{m-1}$ and $E_{m+2}$,
\begin{align}
  \label{eq:energies-soe}
  \begin{pmatrix}
    M_{m-1,m-1} & M_{m-1,m+2} \\
    M_{m+2,m+1} & M_{m+2,m+2}
  \end{pmatrix}
  \begin{pmatrix}
    E_{m-1} \\ E_{m+2}
  \end{pmatrix}
  =
  \begin{pmatrix}
    v_{m-1} \\ v_{m+2}
  \end{pmatrix},
\end{align}
with the matrix entries
\begin{subequations}
  \begin{align}
    M_{m-1,m-1} &= C_{m,m+2} \left( D_{m-1} \jt_{m-1,m} + d C_{m-1,m}
    \right), \\
    M_{m-1,m+2} &= \jt_{m-1,m} \left( D_{m+2} C_{m,m+2} + d
      \jt_{m,m+2} \right), \\
    M_{m+2,m-1} &= D_{m-1} C_{m-1,m+1} \jt_{m+1,m+2} \nn\\
    &\eqph - d \jt_{m-1,m+1} \jt_{m+1,m+2}, \\
    M_{m+2,m+2} &= D_{m+2} C_{m-1,m+1} \jt_{m+1,m+2} \nn\\
    &\eqph - d C_{m-1,m+1} C_{m+1,m+2},
  \end{align}
\end{subequations}
and vector entries
\begin{subequations}
  \begin{align}
    v_{m-1} &= 2 \hbar \dot{\Gamma} n_m - 2 \hbar
    \Gamma j_{m,m+1} + 4 \Gamma^2 n_m \nn\\
    & - D \jt_{m-1,m} C_{m,m+2} - d \jt_{m-1,m} \eta_{m,m+2} \nn\\
    & - d \jt_{m-1,m} \jt_{m,m+2} \left( g_{m+2} n_{m+2} - g_m n_m
    \right) \nn\\
    & - d C_{m-1,m} C_{m,m+2} \left( g_{m-1} n_{m-1} - g_m n_m \right)
    \nn\\
    & + d C_{m,m+2} \zeta_{m+1,m}, \\
    v_{m+2} &= 2 \hbar \dot{\Gamma} n_{m+1} + 2 \hbar \Gamma j_{m,m+1}
    - 4 \Gamma^2 n_{m+1} \nn\\
    & - D C_{m-1,m+1} \jt_{m+1,m+2} - d \jt_{m+1,m+2} \eta_{m-1,m+1}
    \nn\\
    & - d \jt_{m-1,m+1} \jt_{m+1,m+2} \left( g_{m+1} n_{m+1} - g_{m-1}
      n_{m-1} \right) \nn\\
    & - d C_{m-1,m+1} C_{m+1,m+2} \left( g_{m+1} n_{m+1} - g_{m+2}
      n_{m+2} \right) \nn\\
    & + d C_{m-1,m+1} \zeta_{m+1,m+2}.
  \end{align}
\end{subequations}
Note the appearance of the tunneling rates $J_{m-2,m-1}$ and
$J_{m+2,m+3}$. For the four-well system, these quantities do not
appear in the Hamiltonian. In this case, they have to be set to zero
in the equations. If there are more than four wells, the tunneling
rates can be arbitrary and time-dependent. Whereas
Eqs.~\eqref{eq:j-sol} can always be fulfilled, for the linear system
of equations~\eqref{eq:energies-soe} the determinant of the
coefficient matrix must not vanish. This indicates that $\pt$ symmetry
cannot be realized for all conditions and for an arbitrary long time
period. In Sec.~\ref{sec:choices-function-d} we discuss cases in which
the determinant vanishes.

What have we gained by considering the time derivatives of
Eqs.~\eqref{eq:two-cond-n}? We have seen that there are no free
parameters left to fulfill these equations. However, we now have
conditions for the onsite energies $E_{m-1}$ and $E_{m+2}$ such that
the time derivatives of both sides of each equation are equal. If we
make sure that Eqs.~\eqref{eq:two-cond-n} are fulfilled for the
initial time and the onsite energies are chosen according to
Eqs.~\eqref{eq:energies-soe}, then the conditions are fulfilled for
every time. We set the four matrix elements of the
Hamiltonian~\eqref{eq:ham} $J_{m-1,m}$, $J_{m+1,m+2}$, $E_{m-1}$,
$E_{m+2}$, and have one degree of freedom with the function $d$, then
the three independent conditions
\eqref{eq:conditions-1}--\eqref{eq:conditions-3} are fulfilled and we
have a realization of $\pt$-symmetry.

\subsection{Choices for the function \texorpdfstring{$d$}{d}}
\label{sec:choices-function-d}

Before investigating specific scenarios of realizing $\pt$ symmetry,
we discuss two possibilities of choosing the time-dependent function
$d$. We want to point out that a different choice of $d$ does not
change the dynamics of the observables, rather it changes the values
of the matrix elements $J_{m-1,m}$, $J_{m+1,m+2}$, $E_{m-1}$, and
$E_{m+2}$.

\subsubsection{Constant \texorpdfstring{$d$}{d}}
\label{sec:constant-d}

The most simple possibility is a time-independent $d$, i.\,e.\
$\dot{d}=0$, from which it follows that $D_{m-1} = D_{m+2} = D =
0$. Then, the determinant of the coefficient matrix $\mat{M}$ in
Eq.~\eqref{eq:energies-soe} can be expressed with the analytical
results of \cite{kreibich14} as
\begin{align}
  \label{eq:det}
  \det \mat{M} &= \pm 16 d^2 n_{m-1} n_m n_{m+1} n_{m+2} \sqrt{(1 -
    \alpha)^2 - \beta^2}
\end{align}
with the plus or minus sign depending on the initial conditions and
\begin{align}
  \alpha &= (\beta + \gamma) \gamma, & \beta &= \frac{\Gamma}{d
    \sqrt{n_{m-1} n_{m+2}}}, & \gamma &= \frac{\jt_{m,m+1}}{2
    \sqrt{n_m n_{m+1}}}.
\end{align}
When either one of the reservoir or embedded wells is empty, the
conditions cannot be fulfilled. Additionally, the term in the square
root in Eq.~\eqref{eq:det} can be zero, which can be rewritten to
\begin{align}
  \beta = -\gamma \pm 1.
\end{align}
When the number of particles in the reservoir wells, i.\,e.\ $n_{m-1}$
and $n_{m+2}$, is large enough, the linear system of equations can be
solved. However, before these wells get empty, there is the
possibility that there are no solutions for the onsite energies
anymore. The time-independent parameter $d$ can then be changed to
another (larger) value in a new simulation, which does modify the
resulting matrix elements $J_{m-1,m}$, $J_{m+1,m+2}$, $E_{m-1}$, and
$E_{m+2}$.

\subsubsection{Compensating the time-dependence of the tunneling elements}
\label{sec:comp-change-tunn}

Suppose that for the initial time the tunneling elements $J_{m-1,m}$
and $J_{m+1,m+2}$ are given by the trapping geometry. With
Eq.~\eqref{eq:j-sol} these tunneling elements depend on time, which
can be difficult to realize in a given geometry. When we want
$J_{m-1,m}$ to be fixed to the initial value $J_{m-1,m}^{(0)}$, then
we must require
\begin{align}
  J_{m-1,m} = d C_{m,m+2} \stackrel{!}{=} J_{m-1,m}^{(0)}
  \Leftrightarrow d = \frac{J_{m-1,m}^{(0)}}{C_{m,m+2}}.
\end{align}
Similar, if we want $J_{m+1,m+2}$ to be fixed, $d$ must be chosen to
\begin{align}
  d = \frac{J_{m+1,m+2}^{(0)}}{C_{m-1,m+1}}.
\end{align}
Both conditions cannot be fulfilled at the same time, however if we
demand the actual value of $d$ to be the average of both
possibilities, we have an ``average'' fulfillment. The time-dependent
value of $d$ is then
\begin{align}
  \label{eq:d-comp}
  d = \frac{J_{m-1,m}^{(0)}}{2 C_{m,m+2}} + \frac{J_{m+1,m+2}^{(0)}}{2
    C_{m-1,m+1}}.
\end{align}
To calculate the onsite energies, we need the time derivative of
$d$. We get for the parameters in Eq.~\eqref{eq:d-val-def}
\begin{subequations}
  \begin{align}
    D_{m-1} &= \frac{J_{m+1,m+2}^{(0)}}{2 C_{m-1,m+1}^2}
    \jt_{m-1,m+1}, \\
    D_{m+2} &= - \frac{J_{m-1,m}^{(0)}}{2 C_{m,m+2}^2} \jt_{m,m+2},
  \end{align}
  and
  \begin{align}
    D &= - \frac{J_{m-1,m}^{(0)}}{2 C_{m,m+2}^2} \left( g_{m+2}
      n_{m+2} - g_m n_m \right) \jt_{m,m+2} \nn\\
    &\eqph - \frac{J_{m+1,m+2}^{(0)}}{2 C_{m-1,m+1}^2} \left( g_{m+1}
      n_{m+1} - g_{m-1} n_{m-1} \right) \jt_{m-1,m+1} \nn\\
    &\eqph - \frac{J_{m-1,m}^{(0)}}{2 C_{m,m+2}^2} \eta_{m,m+2} -
    \frac{J_{m+1,m+2}^{(0)}}{2 C_{m-1,m+1}^2} \eta_{m-1,m+1}.
  \end{align}
\end{subequations}

As with the case of a constant $d$ we evaluate the determinant of the
coefficient matrix $\mat{M}$ in Eq.~\eqref{eq:energies-soe} to
investigate for which time period the system of equations can be
solved. The determinant can be evaluated analytically, and after a
lengthy calculation we obtain
\begin{multline}
  \label{eq:det2}
  \det \mat{M} = 8 n_{m-1} n_m n_{m+1} n_{m+2} d^2 \\
  \times \left( \pm \sqrt{(1 - \alpha)^2 - \beta^2} + \gamma^2 - 1
  \right).
\end{multline}
Compared to the case of constant $d$ the additional term $\gamma^2-1$
appears in the bracket. As before, when one of the wells is empty the
determinant is zero. For additional roots we have to consider the term
in the bracket. This yields the solutions $\beta=0$ and $\beta = -2
\gamma$ with the plus sign in Eq.~\eqref{eq:det2}, otherwise there are
no solutions. If we can make sure that for a certain realization the
minus sign is chosen in Eq.~\eqref{eq:det2} then the linear system of
equations can be solved as long as the embedded and reservoir wells
are not empty, which is an advantage compared to the case of constant
$d$.

\subsection{Choices for other matrix elements in the multi-mode model}
\label{sec:other-matr-elem}

With the previous investigations we fixed the matrix elements of the
Hamiltonian~\eqref{eq:ham} $J_{m-1,m}$, $J_{m+1,m+2}$, $E_{m-1}$, and
$E_{m+2}$ of the multi-mode model. For the minimal four-mode model
\cite{kreibich13b,kreibich14}, all matrix elements are determined and
no freedom is left, despite the function $d$. When using more than
four wells, the additional tunneling elements and onsite energies are
not fixed by the conditions
\eqref{eq:conditions-1}--\eqref{eq:conditions-3} and may be chosen
arbitrarily. In this section we give the three possibilities that we
use in this work. For all choices we assume that the tunneling
elements are time-independent besides $J_{m-1,m}$ and $J_{m+1,m+2}$,
which are determined by Eqs.~\eqref{eq:j-sol}.

\subsubsection{Leveling out onsite energies}
\label{sec:leveling-out-onsite}

One of the simplest possibilities is to choose the energies of the
wells to the left of the embedded wells equal to $E_{m-1}$, and the
energies of the wells to the right equal to $E_{m+2}$, both of which
are determined by Eqs.~\eqref{eq:energies-soe}, that means
\begin{align}
  E_k=
  \begin{cases}
    E_{m-1}, &\text{for } k<m-1, \\
    E_{m+2}, &\text{for } k>m+2.
  \end{cases}
\end{align}
With this method one effectively enlarges the particle reservoir,
which allows for increasing the time where $\pt$ symmetry is
available. This choice is sketched in Fig.~\ref{fig:basic-idea}(c).

\subsubsection{Requiring specific currents}
\label{sec:requ-spec-curr}

For wells $m-1$ and $m+2$ we have chosen the energies such that the
currents $j_{m-1,m}$ and $j_{m+1,m+2}$ obtain a given value. This
should also be possible for the other currents. We assume target
values of the currents $j_{k,k+1}^\tar$ for $1 \leq k \leq m-2$ and
$m+2 \leq k \leq \numw - 1$ that may be time-dependent. We take the
time derivatives of these currents, which yields with
Eq.~\eqref{eq:jt-deriv}
\begin{multline}
  \label{eq:specific-energies-1}
  E_k = \frac{\hbar^2}{J_{k,k+1} C_{k,k+1}}
  \firsttderiv{j_{k,k+1}^\tar} + \frac{\zeta_{k,k+1}}{C_{k,k+1}} \\
  + E_{k+1} - g_k n_k + g_{k+1} n_{k+1},
\end{multline}
where we consider the case $1 \leq k \leq m-2$ and assume that
$E_{k+1}$ is determined. Beginning with $k=m-2$ we can consecutively
calculate the onsite energies for the wells to the left of the
embedded wells down to the well $k=1$. By similar means we get
\begin{multline}
  \label{eq:specific-energies-2}
  E_{k+1} = - \frac{\hbar^2}{J_{k,k+1} C_{k,k+1}}
  \firsttderiv{j_{k,k+1}^\tar} - \frac{\zeta_{k,k+1}}{C_{k,k+1}} + E_k
  \\
  + g_k n_k - g_{k+1} n_{k+1}.
\end{multline}
for the wells to the right. Note that, as with the onsite energies
$E_{m-1}$ and $E_{m+2}$, the initial wave function $\vec{\psi}$ must
correspond to the target currents, since we only consider the
time derivative to derive the onsite energies. 

\subsubsection{Optical lattice with Stark potential}
\label{sec:extend-optic-latt}

When $E_{m-1}$ and $E_{m+2}$ are given by Eqs.~\eqref{eq:energies-soe}
we can choose the other onsite energies such that they form an optical
lattice with a linear Stark potential with
\begin{align}
  \label{eq:lattice-energies}
  E_k=
  \begin{cases}
    0, &\text{for } k = m, m+1, \\
    \left[ k - \left( m + 1/2 \right) \right] \Delta E + E^{(0)},
    &\text{otherwise},
  \end{cases}
\end{align}
where we have chosen the middle of the embedded wells as the center,
which explains the term $m + 1/2$, see
Fig.~\ref{fig:basic-idea}(d). With $E_{m-1}$ and $E_{m+2}$ the slope
$\Delta E$ and offset $E^{(0)}$ can be determined, which yields
\begin{subequations}
  \begin{align}
    \Delta E &= \frac{1}{3} \left( E_{m+2} - E_{m-1} \right), \\
    E^{(0)} &= \frac{1}{2} \left( E_{m-1} + E_{m+2} \right).
  \end{align}
\end{subequations}
Note that $E_m = E_{m-1} = 0$ are not determined by the linear Stark
potential so that they can be interpreted as a perturbation. This
possibility should allow for an easy experimental realization as an
optical lattice with a linear Stark potential is an approved technique
used in many experiments.

\subsection{Fourier ansatz for an optical lattice}
\label{sec:feat-optic-latt}

The matrix model given by the Hamiltonian~\eqref{eq:ham} can be solved
by the usual methods: Either one can diagonalize the Hamiltonian
matrix analytically or by using numerical methods. However, for the
special case of an optical lattice with an infinite number of wells,
where the onsite energies are the same for each well, there are
specific solution methods available. A simple quantum mechanical model
for an optical lattice is the discrete Schrödinger equation with an
infinite number of wells, where the Hamiltonian is defined as
\begin{align}
  \label{eq:ham-optical-lat}
  H_{kl} =
  \begin{cases}
    -J, & \text{for } \abs{k-l} = 1, \\
    0, & \text{else}.
  \end{cases}
\end{align}
The onsite energies are set to zero, and the tunneling elements are
assumed to be the same for each well barrier. The eigenstates are
given by
\begin{align}
  \label{eq:lattice-sol}
  \psi_{q,n} = A \exp \left( \ii q n \right)
\end{align}
with lattice index $n$ and quantum number $q$, with the eigenvalues
\begin{align}
  E(q) = -2 J \cos q,
\end{align}
where $q \in [-\pi,\pi]$. With these results it is clear that our
model features only one band as opposed to a more extended
model~\cite{morsch06}, where the continuous Schrödinger equation is
solved. When we include an interaction in our model via the mean-field
approximation with non-vanishing interaction strengths $g_k$ the
eigenstates~\eqref{eq:lattice-sol} are still solutions. However, new
solutions appear that may break the translational symmetry of the
model, and the solutions~\eqref{eq:lattice-sol} may become unstable
with respect to small perturbations, both of which we will not
consider in this work.

A general time-dependent solution of the linear Schrödinger equation
has the form
\begin{align}
  \label{eq:lattice-gen-sol}
  \psi_n(t) = \int\limits_{-\pi}^\pi \dd q \, f(q) \exp \left[ \ii
    \left( q n - E(q) t / \hbar \right) \right].
\end{align}
An analytical solution of this equation is not available. However,
there is an important approximation called the \emph{semiclassical
  approximation} that is valid, when the amplitude $f(q)$ is sharply
centered around a value $q_0$. In this case the wave function reads
\cite{morsch06}
\begin{multline}
  \label{eq:lattice-semicals}
  \psi_n(t) = \exp \left[ \ii \left( q_0 n - E(q_0) t / \hbar \right)
  \right] \\
  \times \int\limits_{-\pi}^\pi \dd q \, f(q) \exp \left[ \ii \left( n
      - v_\group(q_0) t \right) \left( q - q_0 \right) \right] \\
  \times \exp \left[ - \ii \hbar t \left( q - q_0 \right)^2 / 2
    m_\eff(q_0) \right]
\end{multline}
with
\begin{align}
  v_\group(q_0) &= E^\prime(q_0) / \hbar, & m_\eff(q_0) &= \hbar^2 /
  E^{\prime\prime}(q_0)
\end{align}
the \emph{group velocity} and \emph{effective mass}, respectively. The
wave packet moves with group velocity $v_\group$ and disperses
analogously to a free wave packet (without a potential) describing a
particle with mass $m_\eff$. Connected with the propagating wave
packet is a current: When a wave packet moves, for instance, to the
right and a single well is considered, there is a particle flow to
this well out of the left side and a flow from this well to the right
side.

With the Hamiltonian given by Eq.~\eqref{eq:ham-optical-lat} the
movement of a wave packet cannot be changed. We therefore introduce a
linear Stark potential, where the onsite energies are given by $E_k =
\Delta E k$. When $\Delta E$ is constant and the semiclassical
approximation is used, the center of the momentum distribution gets
time-dependent according to $q_0(t) = q_0(0) + \Delta E t / \hbar$
\cite{morsch06}. The quasi-momentum $q_0$ has a linear
time-dependence, but when inserted into the group velocity, this leads
to
\begin{align}
  v_\group(q_0) = E^\prime(q_0) / \hbar = \frac{2 J}{\hbar} \sin
  \left[ q_0(0) + \frac{\Delta E t}{\hbar} \right].
\end{align}
The group velocity oscillates in time, and with it the position of the
wave packet according to $\dot{n}_0 = v_\group(q_0)$. This is an
occurrence of the famous \emph{Bloch oscillation} due to a constant
force in a periodic potential. This model can be extended to a
time-dependent force $\Delta E$ by
\begin{align}
  \label{eq:quasi-momentum}
  \dot{q}_0(t) &= \Delta E(t) / \hbar, & \dot{n}_0(t) =
  v_\group(q_0(t)).
\end{align}
Now the time derivative of the pseudo-momentum $q_0$ is proportional
to the force $\Delta E$, which explains the notion ``semiclassical
approximation''. The term in Eq.~\eqref{eq:lattice-semicals}
containing the effective mass must also be generalized according to
\begin{align}
  \frac{t}{m_\eff(q_0)} \to \int\limits_0^t \dd t^\prime \,
  \frac{1}{m_\eff(q_0(t^\prime))} \equiv \frac{t}{\bar{m}_\eff(t)}.
\end{align}

When the sharp momentum peak of $f(q)$ in
Eq.~\eqref{eq:lattice-semicals} does not come close to the border of
the Brillouin zone, we can extend the interval of the integral in
Eq.~\eqref{eq:lattice-semicals} to $[-\infty, \infty]$ and we are able
to explicitly calculate the wave function. We make use of this
possibility when discussing the realization of $\pt$-symmetry in terms
of optical lattices in Sec.~\ref{sec:optical-lattices}.

\section{Results of simulations}
\label{sec:results-simulations}

In Sec.~\ref{sec:real-pt-symm} we developed the central idea to
realize $\pt$ symmetry by embedding a $\pt$-symmetric two-well system
into a larger Hermitian system. It is the purpose of this section to
apply the theoretical results and to investigate numerically actual
realizations. The most simple system with which we can realize a
$\pt$-symmetric system contains at each side one additional reservoir
well, which gives in total a four-well system. Results for this
minimal four-well system are presented in our Rapid
Communication~\cite{kreibich13b} and more comprehensively
in~\cite{kreibich14}. We showed that with a four-well system we can
realize $\pt$-symmetry for suitable long times and that by choosing
the reservoir wells larger and larger we can extend the time that is
available for realizing $\pt$-symmetry. This, however, is in an
experimental setup limited for two reasons. At first, a large
difference in the particle population between the inner and outer
wells can be difficult to measure in an experiment. In addition, a
large population in the reservoir wells can make the onsite energies
extremely large, which would be another difficulty in an experimental
setup.

For this reasons we continue the investigations started in
\cite{kreibich13b} by choosing more than two reservoir wells and
consequently increasing the number of wells which leads us to the
discussion of $\pt$-symmetric currents in optical lattices with a
linear Stark potential. Here, we give only a few example
calculations. More can be found in \cite{kreibich-diss}.

\subsection{Six-mode model}
\label{sec:six-mode-model}

\subsubsection{Leveling out onsite energies}
\label{sec:leveling-out-onsite-1}

We start with the six-mode model with one additional reservoir well on
each side compared to the four-mode model. This is the most simple
extension of the four-mode system with a symmetric number of wells on
each side of the embedded wells. It effectively enlarges the
reservoir, since the reservoir wells directly coupled to the embedded
wells can now be fed by the outer wells and thus the time that is
available for realizing $\pt$ symmetry is extended. As a first example
we consider the case where the additional wells are chosen to be at
the same energy level as the wells directly coupled to the embedded
wells, so called \emph{leveled out} energies. For the six-mode model
this implies
\begin{align}
  E_1 &= E_2, & E_5 &= E_6,
\end{align}
where wells $m=3$ and $m+1=4$ are the embedded wells. All matrix
elements are now fixed, solely the initial wave functions in wells $1$
and $6$ are arbitrary and thus the initial populations and currents to
the adjacent wells. A somewhat naive approach to choose arbitrary
values of the parameters of the initial wave function is a
possibility, but this results in large oscillations between the
reservoir wells, which is a valid, but not a suitable realization for
an experiment.

\begin{figure}[t]
  \centering
  \includegraphics[width=\linewidth]{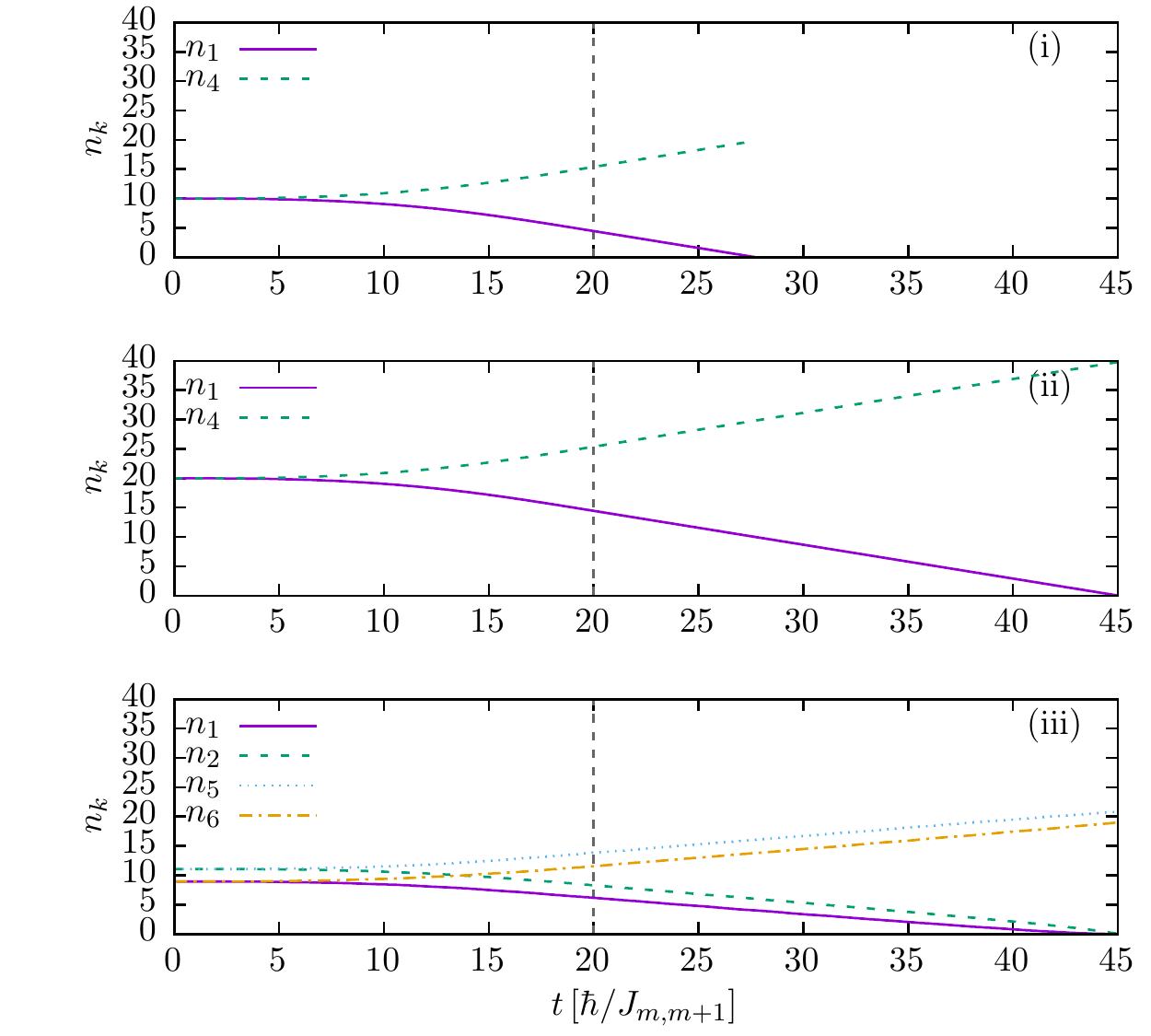}
  \caption{(Color online) (i) Population of the outer wells for the
    four-mode model, where an adiabatic current ramp is simulated with
    a final time of $t_\tar = 20 \hbar / J_{m,m+1}$. (ii) The time
    that is available for realizing $\pt$ symmetry should be
    extended. On the one hand one can simply increase the initial
    number of particles in the reservoir wells. (iii) On the other
    hand the six-mode model can be used to effectively enlarge the
    reservoir. Now the difference of the particle numbers between the
    embedded and reservoir wells is smaller, which could help to
    measure the well population in an experiment. The nearly constant
    population of the embedded wells $n_m(t) = n_{m+1}(t) \approx 0.5$
    is not shown.}
  \label{fig:level-probs}
\end{figure}

\begin{figure}[t]
  \centering
  \includegraphics[width=\linewidth]{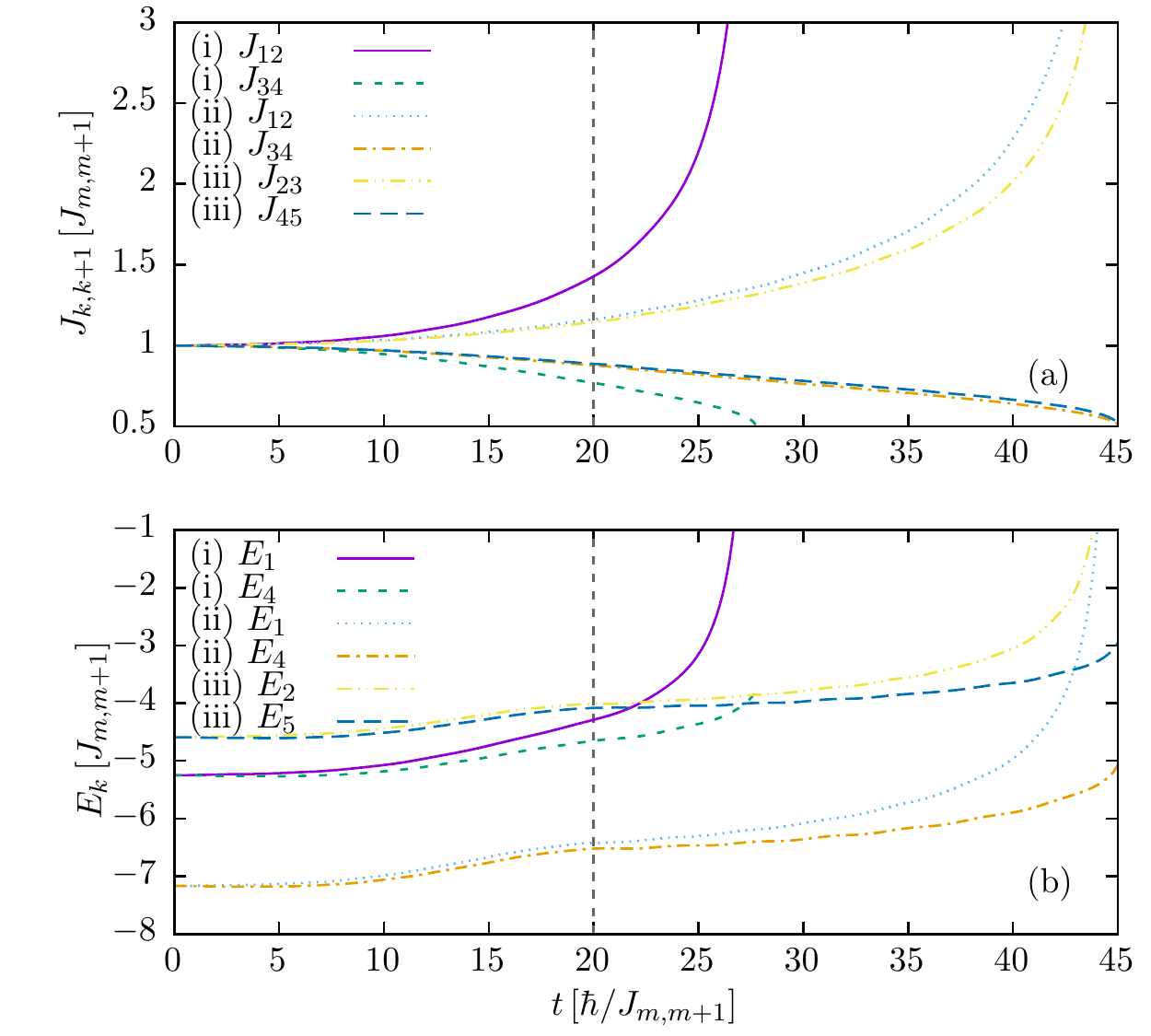}
  \caption{(Color online) (a) Tunneling elements $J_{m-1,m}$,
    $J_{m+1,m+2}$ and (b) onsite energies $E_m$, $E_{m+1}$ for the
    scenarios (i)--(iii) discussed in
    Fig.~\ref{fig:level-probs}. Comparing (ii) and (iii), for the
    tunneling elements there is not much difference. The onsite
    energies, however, differ by some amount. The magnitude for the
    six-mode model is smaller, which could be an advantage in an
    experiment.}
  \label{fig:level-elements}
\end{figure}

Hence, the initial conditions must be chosen carefully. A natural
solution is the use of the adiabatic current ramp \cite{kreibich14},
where the initial condition is the ground state of a Hermitian system
and the $\pt$-parameter is time-dependently varied as
\begin{align}
  \Gamma(t) =
  \begin{cases}
    0, & \text{for } t < 0, \\
    \Gamma_\tar \left[ 1 - \cos \left( \pi t / t_\tar \right) \right]
    / 2 & \text{for } 0 \leq t \leq t_\final, \\
    \Gamma_\tar, & \text{for } t > t_\tar
  \end{cases}
\end{align}
with target parameter $\Gamma_\tar$ reached at $t = t_\tar$. To
investigate the use of the six-mode model, we simulate three different
scenarios that allow us a comparison. All cases have the adiabatic
current ramp with a target value of $\Gamma_\tar / J_{m,m+1} = 0.5$ in
common, which is reached at a target time of $t_\tar = 20 \hbar /
J_{m,m+1}$. The initial population for the embedded wells is always
$n_m = n_{m+1} = 0.5$. (i) For the first run, we use the four-mode
model with an initial population for the reservoir wells of $n_1 = n_4
= 10$. (ii) Still with the four-mode model, for the second simulation
the particle number in the reservoir wells is doubled to $n_1 = n_4 =
20$. (iii) For the third scenario we use the six-mode model, where the
onsite energies are chosen such that for the initial time we have $n_1
+ n_2 = n_5 + n_6 = 20$, giving the same number of reservoir particles
as in the second case. Figure~\ref{fig:level-probs} shows the
time-dependent populations of the wells, where a non-interacting
\ac{BEC} is simulated. The first two cases are analogous to the
results of \cite{kreibich14}. For the six-mode model we see that the
reservoir wells left of the embedded wells are emptying with a slower
speed, which was expected, since both wells serve as a common
reservoir. The wells to the right are filling analogously. The time
that is available for realizing $\pt$-symmetry is the same for the
last two cases, hence the six-mode model gives no advantage with
respect to the time that is available. However, there are other
advantages discussed in the following.

Of further interest is the time-dependence of the matrix elements,
which is shown in Fig.~\ref{fig:level-elements}. The tunneling
elements show almost the same behavior for the cases~(ii)
and~(iii). The difference lies in the onsite energies. Due to the
smaller amount of particles per well for the six-mode model the
magnitude of the onsite energies is smaller than in case~(ii), as well
as the rate of change. This could result in an easier experimental
realization.

With more wells we now have a second possibility to enlarge the
particle reservoir. Besides putting more particles in the two outer
wells of the four-mode model we can add more wells. The difference is
the resulting onsite energies, which with this method can be brought
into an experimentally accessible range.

\subsubsection{Requiring specific currents}
\label{sec:requ-spec-curr-1}

The second possibility for choosing the onsite energies is by
requiring specific currents between the additional wells. Of course,
the choice of these target currents is free. A somewhat natural choice
is to require that each of the reservoir wells should contribute
equally to the $\pt$-symmetric currents. For our six-mode model this
means that when wells $3$ and $4$ are the embedded wells, the current
$j_{23}$ is given by the condition~\eqref{eq:conditions-1}, and for
$j_{12}$ we require $j_{12} = j_{23}/2$. Similarly we set $j_{56} =
j_{45}/2$. The reservoir wells on the left should then be emptying at
the same speed, as well as those on the right should fill. The initial
conditions and the choice of the onsite energies are then given by the
results in Sec.~\ref{sec:other-matr-elem}. This leads to a scenario
that is similar to that of leveling out the onsite energies, the
difference is that now we can choose arbitrary initial conditions for
the embedded wells and the phases of the reservoir wells can be
calculated appropriately, thus extending the possibilities.

When the reservoir wells to the left are emptying at the same speed,
this can be unfavorable when one well has a lower initial population
than the other: When this well gets empty, the simulation stops
although there are particles in the left reservoir. This can be
overcome by choosing the specific inter-well currents proportional to
the initial population. Suppose that $j_{m-1,m}$ and $j_{m+1,m+2}$ are
given by conditions~\eqref{eq:conditions}. Then we set for the
reservoir wells $\dot{n}_k^\tar = j_{k-1,m}^\tar - j_{k,k+1}^\tar
\propto n_k(t=0)$. The proportionality constant can be fixed by
requiring $j_{m-1,m}$ and $j_{m+1,m+2}$ to fulfill the conditions. In
the six-mode model we get for the left reservoir wells
\begin{align}
  j_{12}^\tar &= - C n_1(0), & j_{12}^\tar - j_{23}^\tar &= C n_2(0).
\end{align}
This can be solved according to
\begin{align}
  j_{12}^\tar &= - C n_1(0), & j_{23}^\tar &= - C \left[ n_1(0) +
    n_2(0) \right].
\end{align}
For $j_{23}$ we get from condition~\eqref{eq:conditions-1}
$j_{23}^\tar = 2 \Gamma n_3(0) / \hbar$. The constant $C$ can now be
determined and we obtain
\begin{align}
  j_{12}^\tar &= \frac{2 \Gamma n_1(0) n_3(0)}{\hbar \left[ n_1(0) +
      n_2(0) \right]}, & j_{23}^\tar &= \frac{2 \Gamma n_3(0)}{\hbar}.
\end{align}
For an arbitrary number of wells, this can be solved by iteration. The
target currents for the right wells are not essential, hence we can
choose them to be symmetric to the left wells.

\begin{figure}[t]
  \centering
  \includegraphics[width=\linewidth]{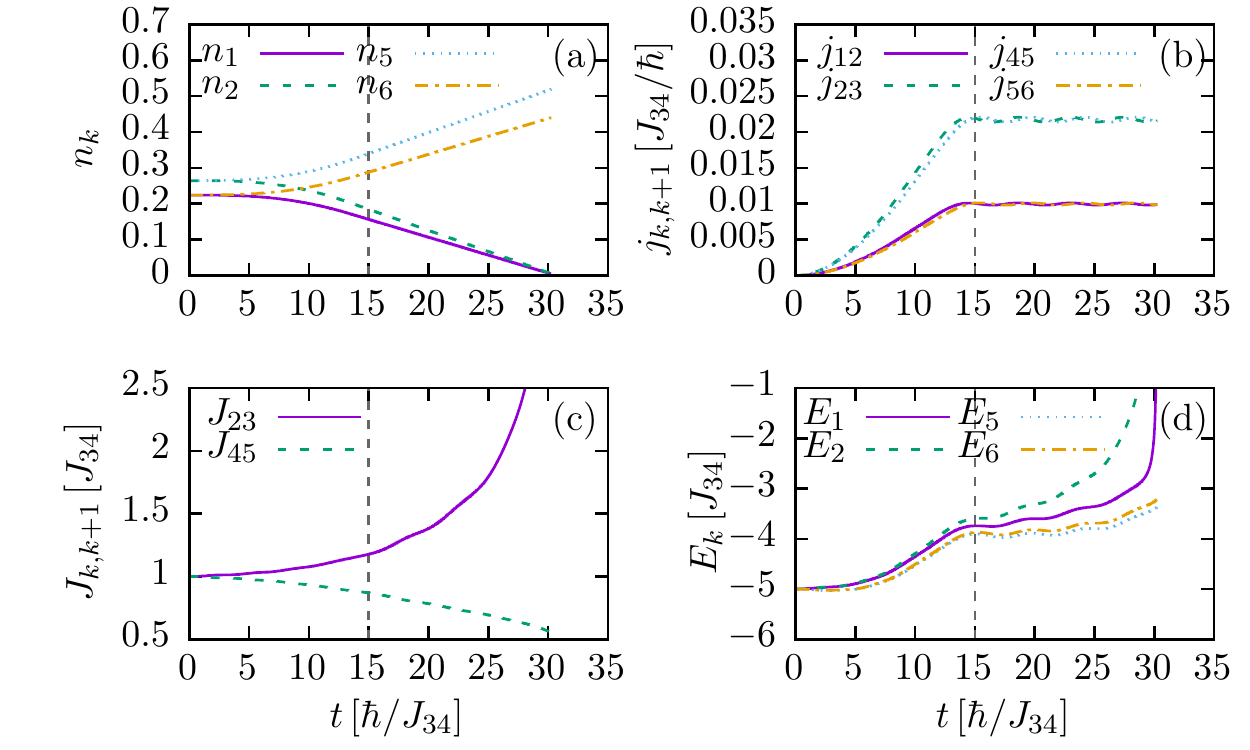}
  \caption{(Color online) Adiabatic current ramp with two reservoir
    wells on each side with $\numw=6$ wells. Shown are (a) particle
    numbers in the reservoir wells, (b) inter-well currents, (c)
    tunneling elements, and (d) onsite energies. We start with the
    ground state for the parameters given in the text. The target
    currents are chosen in such a way that the left reservoir wells
    get empty at nearly the same time. This allows us to effectively
    use all particles in the reservoir wells and thus to maximize the
    time that is available for $\pt$-symmetry.}
  \label{fig:specific_prop}
\end{figure}

We apply this scenario to a system, where we start with the ground
state for $E_k / J _{34} = -5$ for $k = 1,2,5,6$, $E_k = 0$ for
$k=3,4$, $J_{k,k+1} / J_{34} = 1$ and $g = 1$. In this case, the
ground state is normalized to unity, $\abs{\vec{\psi}}^2 = 1$. We use
the adiabatic current ramp with $\Gamma_\final / J_{34} = 0.7$ and
$t_\final = 15 \hbar / J_{34}$. The observables and matrix elements
are shown in Fig.~\ref{fig:specific_prop}. Clearly, the initial
population of wells $1$ and $2$ are different, however, they get empty
at approximately the same time. The target currents obey $j_{12} /
j_{23} = j_{56} / j_{45} \approx 0.4589$. With this choice of the
currents the onsite energies are no longer approximately equal, so
this scenario differs from the previous one. We can now start from
arbitrary initial conditions and realize $\pt$ symmetry with more than
one reservoir well on each side until all left reservoir wells become
empty. Here we give a specific example of an adiabatic current ramp,
however, with this choice of target currents we can simulate arbitrary
scenarios, that means, starting with a value of $\Gamma \neq 0$ and
non-quasi-stationary solutions.

\subsection{More wells}
\label{sec:more-wells}

When using more than six wells, as in Sec.~\ref{sec:six-mode-model},
the options are consequently extended. In this section, we give a
single example of what can be done.

\begin{figure}[t]
  \centering
  \includegraphics[width=\linewidth]{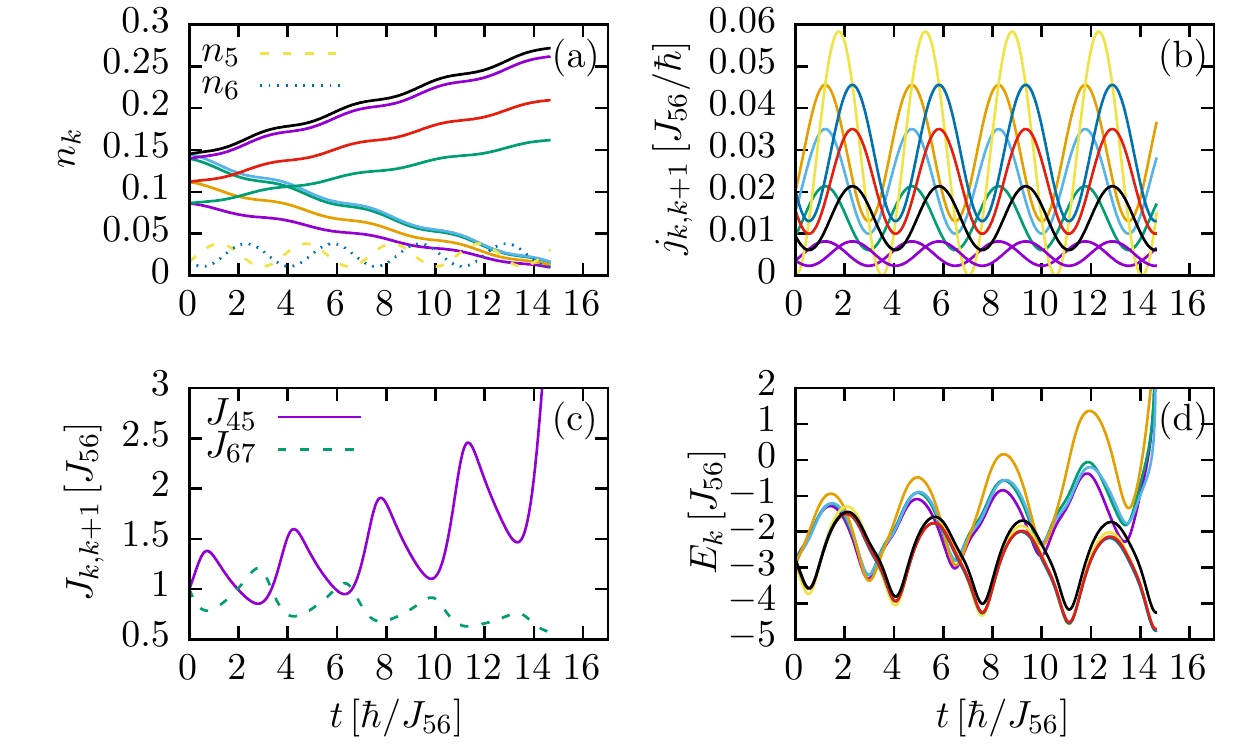}
  \caption{(Color online) Non-quasi-stationary solutions for the
    parameters $\Gamma / J_{56} = 0.6$ and $g / J_{56} = 10$ with
    $\numw=10$ wells. See text for initial conditions. Shown are (a)
    particle numbers, (b) inter-well currents, (c) tunneling elements,
    and (d) onsite energies. The currents are chosen such that the
    reservoir wells are emptying with a rate proportional to the
    initial population. With this choice the reservoir can be used in
    an optimal way. Here we give up the leveled out onsite energies.}
  \label{fig:specific_dynamics_10}
\end{figure}

We realize $\pt$ symmetry using $N_\well = 10$ wells and require the
currents in such a way that the contribution from each reservoir well
is proportional to its initial population. We set $\Gamma / J_{56} =
0.6$, $g / J_{56} = 10$, and start with the ground state for $E_k = 0$
for $k = 5, 6$ and $E_k / J_{56} = -3$ otherwise, furthermore we set
for all $k$ $J_{k,k+1} / J_{56} = 1$. Since we start with a finite
value of $\Gamma$, where the ground state matches to $\Gamma = 0$,
this corresponds to a non-quasi-stationary solution. The observables
and matrix elements are shown in
Fig.~\ref{fig:specific_dynamics_10}. The number of particles in the
reservoir well on the left of the embedded wells decreases in such a
way that the wells get empty at the same time. Thus, the whole
reservoir is used. The currents on the left side are proportional to
each other, as are those on the right side. The energies on each side
are no longer equal, which shows that leveling out the onsite energies
does not use the reservoir in an optimal way.

\subsection{Optical lattices}
\label{sec:optical-lattices}

The use of more and more wells motivates the investigation of the
realization of $\pt$-symmetric \acp{BEC} in terms of optical lattices,
of which the basics are discussed in
Sec.~\ref{sec:feat-optic-latt}. Instead of requiring a specific
current from each reservoir well, we now simply calculate the energy
values of the wells adjacent to the embedded wells and extend the
remaining onsite energies such that they build an optical lattice with
a linear Stark potential. This idea has been theoretically worked out
in Sec.~\ref{sec:other-matr-elem}.

In our realization, we always consider a \emph{finite} number of
wells, as opposed to an infinitely extended optical lattice. Instead
of a continuous quantum number $q$ we have a discrete variable, and we
must replace the integral in Eq.~\eqref{eq:lattice-gen-sol} according
to
\begin{align}
  \int\limits_{-\pi}^\pi \dd q \to \sqrt{\frac{2 \pi}{L}}
  \sum\limits_q,
\end{align}
where $L$ is the extension of the finite lattice. The allowed values
of $q$ are then $q = 2 \pi m / L$ with $m = -(L-1)/2, -(L-1)/2+1,
\dots, (L-1)/2$. However, when there is a large number of wells, the
use of a continuous quantum number is a valid approximation. We can
therefore use the semiclassical approximation discussed in
Sec.~\ref{sec:feat-optic-latt} as a description of our system with a
time-dependent force $\Delta E(t)$ to manipulate the wave
packet. However, a pure Gaussian wave packet cannot describe a
$\pt$-symmetric situation since the population in the embedded wells
cannot be constant. Hence, a suitable perturbation of the linear Stark
potential must be added.

In Sec.~\ref{sec:other-matr-elem} we discussed how the energies of the
wells adjacent to the embedded wells can be extended to an optical
lattice with a linear Stark potential according to
Eq.~\eqref{eq:lattice-energies} with a slope $\Delta E$ of the linear
potential and an offset $E^{(0)}$. The offset, where the embedded
wells are \emph{not} part of the linear potential, is the perturbation
that is necessary to obtain a $\pt$-symmetric realization as discussed
in Sec.~\ref{sec:feat-optic-latt}. With this choice of the onsite
energies a realization of $\pt$ symmetry is possible in our developed
framework and the formalism of Sec.~\ref{sec:feat-optic-latt} can be
used to interpret the results.

\begin{figure}[t]
  \centering
  \includegraphics[width=\linewidth]{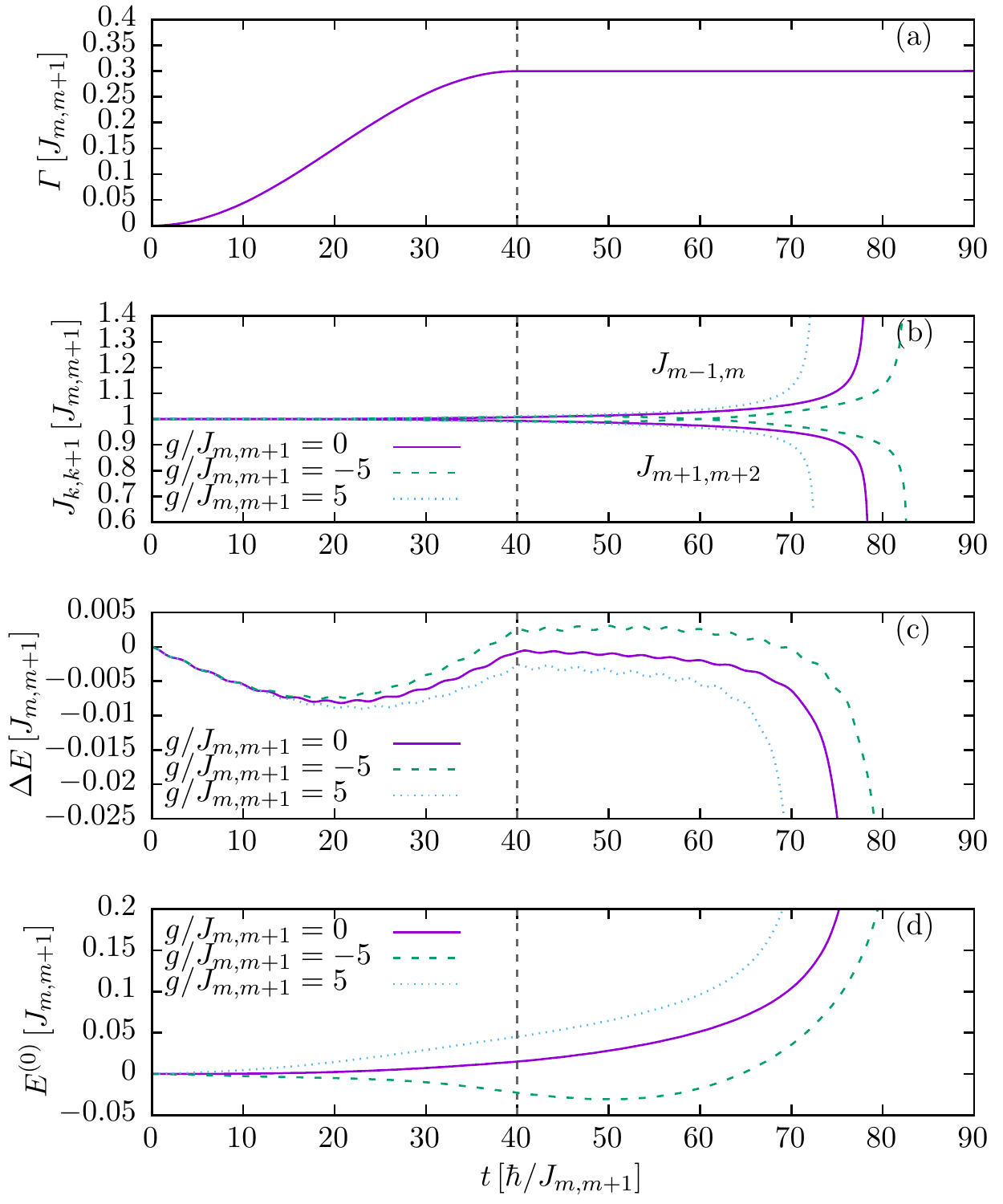}
  \caption{(Color online) (a) $\pt$ parameter $\Gamma$, (b) tunneling
    elements, (c) energy slop $\Delta E$, and (d) energy offset
    $E^{(0)}$ of the energy
    distribution~\eqref{eq:lattice-energies}. For $t = 0$ a Gaussian
    wave function has been prepared and the $\pt$ symmetry is realized
    using an adiabatic current ramp for $\Gamma_\final / J_{m,m+1} =
    0.3$ and $t_\final = 40 \hbar / J_{m,m+1}$. The calculation is
    repeated for different values of the interaction strength. For $t
    < t_\final$ the slope $\Delta E$ is used to accelerate the wave
    packet, whereas for $t > t_\final$ the slope must be changed to
    compensate the decrease in population of the moving wave
    packet. The offset $E^{(0)}$ is the necessary perturbation so that
    $\pt$ symmetry can be realized.}
  \label{fig:energies}
\end{figure}

\begin{figure}[t]
  \centering
  \includegraphics[width=\linewidth]{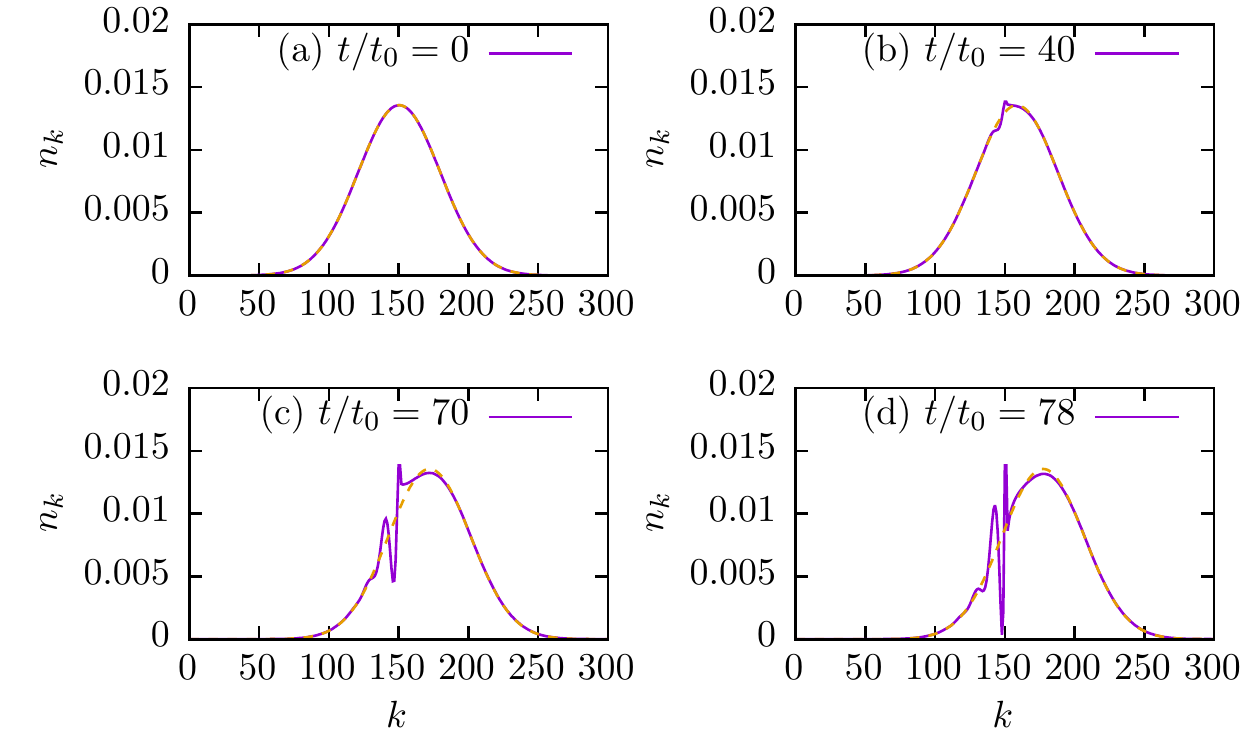}
  \caption{(Color online) Wave function in position space of the
    scenario of Fig.~\ref{fig:energies} for different times (a)--(d)
    and a vanishing interaction strength. The initial Gaussian wave
    packet is accelerated to the right due to the negative energy
    slope $\Delta E$. At the same time the purely Gaussian shape is
    perturbed due to the energy offset. In combination, these effects
    contribute to the realization of $\pt$-symmetry in the embedded
    wells $m=150$ and $m+1=151$ with a constant number of particles
    $n_{150}$ and $n_{151}$. The dashed curves give the wave function
    of the semiclassical approximation, where the energy offset is
    ignored.}
  \label{fig:lattice-wavefunction}
\end{figure}

For the following simulations we use $N_\well = 300$ wells, where the
embedded wells are given by $m=150$ and $m+1=151$, and an adiabatic
current ramp with $\Gamma_\final / J_{m,m+1} = 0.3$ and $t_\final = 40
\hbar / J_{m,m+1}$. The starting wave function is given by
\begin{align}
  \label{eq:lattice_wf_result}
  \psi_n(t=0) = \sqrt[4]{\frac{2 (\Delta q)^2}{\pi}} \exp \left[ -
    (\Delta q)^2 (n - n_0(t=0))^2 \right]
\end{align}
with the parameters $\Delta q = 0.017$, and $n_0(t=0) = 150.5$ such
that the wave function has no momentum at the beginning and fulfills
$n_m = n_{m+1}$. The quantity $\Delta q$ is chosen so that there are
no boundary effects during the simulation
time. Figure~\ref{fig:energies} shows the tunneling elements as well
as the parameters $\Delta E$ and $E^{(0)}$ for different strengths of
interaction. The tunneling elements do not change much over a large
range of the simulation time, see Fig.~\ref{fig:energies}(b). Only
close to the point, where the realization is no longer possible, the
tunneling elements vary noticeably. The time-dependence of the slope
$\Delta E$ shown in Fig.~\ref{fig:energies}(c) can be explained as
follows: At the beginning the $\pt$-symmetric current is being
increased (see Fig.~\ref{fig:energies}(a)) by a negative slope of the
linear energy part, and the wave packet is accelerated to the
right. At the end of the adiabatic current ramp $f_\final = 40 \hbar /
J_{m,m+1}$, the slope is again close to zero, the acceleration stops
and there is a constant current. When the wave packet now moves to the
right (as shown in Fig.~\ref{fig:lattice-wavefunction}), the
population in the reservoir well left of the embedded wells
decreases. This is compensated by an increasing negative slope. When
coming close to the end of the simulation time, the negative slope
must increase and finally diverges to $-\infty$. The influence of the
interaction strength can be understood by noticing that a repulsive
interaction leads to a faster dispersion of the wave packet and hence
a slightly larger negative slope is necessary. The opposite is true
for an attractive interaction, for some time interval the slope even
can get positive. The energy offset $E^{(0)}$ is zero at the
beginning, and then must change to maintain $\pt$ symmetry. The more
distant the center of the wave packet has moved, the more the offset
must vary, which finally leads to a divergence of $E^{(0)} \to
\infty$.

Figure~\ref{fig:lattice-wavefunction} shows the population per well
for different times. The initial Gaussian wave packet in
Fig.~\ref{fig:lattice-wavefunction}(a) is soon disturbed by the energy
offset $E^{(0)}$. Close to the left of the embedded wells is a dip in
the population which further decreases during the simulation. The wave
function is highly perturbed close to the embedded wells, however, for
the wells far away the wave packet is well described by the
semiclassical approximation. Close to the time when the simulation
breaks down particles are still left in the reservoir wells. We
conclude that with the energy distribution according to
Eq.~\eqref{eq:lattice-energies} the particles in the reservoir cannot
be completely used, which is the cost of this simple energy
distribution.

\begin{figure}[t]
  \centering
  \includegraphics[width=\linewidth]{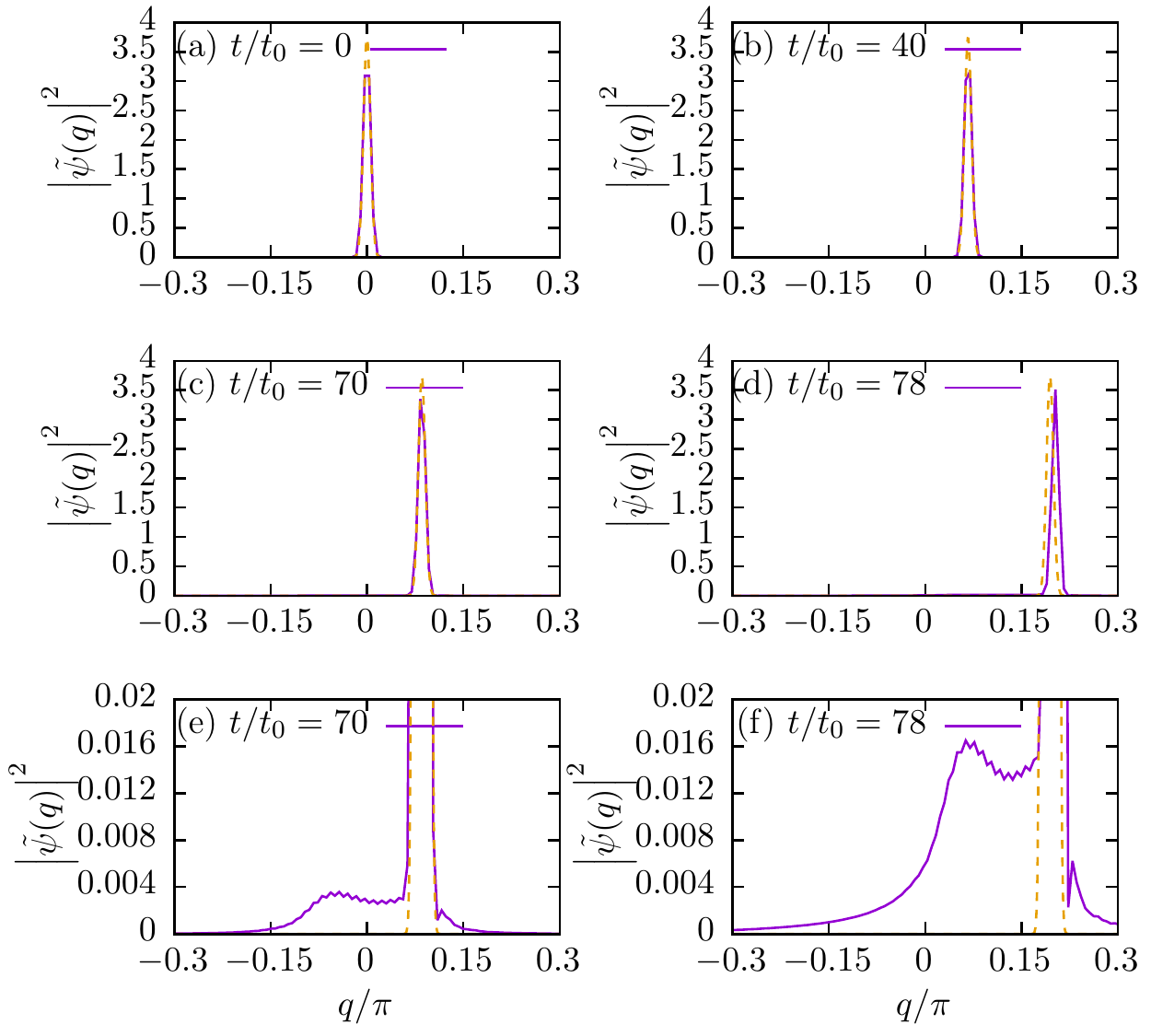}
  \caption{(Color online) Same as Fig.~\ref{fig:lattice-wavefunction},
    but for the wave function in momentum space. The dashed line gives
    the wave function of the semiclassical approximation. The last row
    (e)--(f) shows an enlarged plot of the previous row (c)--(d). In
    momentum space, the wave function is dominated by the sharp peak
    in $q_0$, of which the time-dependence can be calculated according
    to the semiclassical approximation and which gives a good
    description for the dominating peak. Only in the enlarged views
    (e)--(f) one notices the perturbation due to the energy
    offset. This perturbation is getting stronger during the
    simulation time, however, it does not reach the magnitude of the
    momentum peak.}
  \label{fig:lattice-wavefunction-k}
\end{figure}

Of further interest is the Fourier transform of the wave function that
can be calculated according to
\begin{align}
  \tilde{\psi}(q) = \frac{1}{2 \pi} \sum\limits_k \exp \left( -\ii q k
  \right) \psi_k.
\end{align}
Figure~\ref{fig:lattice-wavefunction-k} shows this Fourier transform
for a selection of different times. As to be expected, it is dominated
by a sharp peak in $q$-space. This dominance does not vanish during
simulation time and can be reasonably described by the semiclassical
approximation. The position of this peak moves according to
Eq.~\eqref{eq:quasi-momentum}. The energy offset induces a
perturbation, which is in particular visible near the end of the
simulation time $t \gtrsim 70 \hbar/J_{m,m+1}$. Note that its maximum
magnitude is smaller than one percent of the momentum peak.

\section{Conclusion and outlook}
\label{sec:conclusion-outlook}

We have developed the basic idea to realize a $\pt$-symmetric two-mode
model by embedding this system into a larger Hermitian system with in
total at least four wells where the additional wells are considered as
reservoir wells. In the limit of large number of wells $\pt$ symmetry
is realized by a tilted optical lattice where the embedded double well
is a necessary defect. The coupling of the embedded wells to the
reservoir wells is due to the overlapping wave functions between
adjacent wells. It has already been shown that the necessary potential
can be created experimentally \cite{henderson09}. The coupling of
overlapping wave functions is experimentally accessible contrary to
the methods used in \cite{single14,gutoehrlein15}, where a coupling is
assumed without giving possible realizations of such a coupling and
thus leaving some ambiguity. With our method we are thus able to give
a realistic scenario for realizing $\pt$ symmetry where all parameters
used in a possible experiment are determined.

For the theoretical investigations we used the matrix model developed
in \cite{kreibich14}, where we concentrated on using the four-mode
model.  We applied all the theoretical results to calculate the
dynamical evolution of the realizations of $\pt$ symmetry. Due to a
finite particle reservoir the time that is available for realizing
$\pt$ symmetry is always limited. For a four-well system this time can
be extended by filling more particles in the reservoir wells. However,
a large difference between the number of particles in the embedded
wells on one hand and the particle population in the embedded wells on
the other hand can be a huge experimental challenge for both preparing
such a state and measuring the particle numbers. Furthermore, such a
large difference can make the necessary matrix elements of the
Hamiltonian~\eqref{eq:ham} extraordinarily large. For these reasons,
we extended the model to more than two reservoir wells in this
article, such that there are more possible realizations with an
arbitrary number of reservoir wells. Thus, we have a wealth of
possible realization of $\pt$ symmetry, and an experimentalist can
choose the system parameters such that they are in an accessible
range. We analyzed a system with a large number of wells in the
framework of optical lattices. In that sense, $\pt$ symmetry can be
interpreted by a moving and accelerating wave packet that scatters at
a defect of a tilted optical lattice, which leads to a realization of
$\pt$ symmetry at that double well.

Many of the investigations of $\pt$-symmetric \acp{BEC} were performed
in terms of the mean-field approximation. First extensions were made
\cite{dast14}, and it could be possible to study a realization using
the many-particle Bose-Hubbard model. The results from the matrix
model in this work could then be used as a possible road map. Due to
the higher degrees of internal freedom a comparison with the results
obtained from the many-particle quantum master equation would be of
special interest.

\begin{acknowledgments}
  This work was supported by DFG\@. M.\,K.\ is grateful for support
  from the Landesgraduiertenf\"orderung of the Land
  Baden-W\"urttemberg.
\end{acknowledgments}

%

\end{document}